# Reconstructing Past Solar Activity using Meridian Solar Observations: the Case of the Royal Observatory of the Spanish Navy (1833–1840)

J.M. Vaquero[1*] and M.C. Gallego[2]

[1]Departamento de Física, Centro Universitario de Mérida, Universidad de Extremadura, Mérida, Spain

[2]Departamento de Física, Facultad de Ciencias, Universidad de Extremadura, Badajoz, Spain

**Abstract**

Solar meridian observations have been used to evaluate the solar activity of the past. Some important examples are the solar meridian observations made at the Basilica of San Petronio in Bologna by several astronomers and the observations made by Hevelius published in his book *Machina Coelestis*. However, we do not know whether these observations, which were not aimed to estimate the solar activity, are reliable for evaluating solar activity. In this paper, we present the marginal notes about sunspots that are included in the manuscripts of the meridian solar observations made at the Royal Observatory of the Spanish Navy during the period 1833–1840. We compare these observations with other solar activity indices such as sunspot area and number. Our conclusion is that solar meridian observations should be used with extreme caution to evaluate past solar activity.

**1. Introduction**

In the last fifteen years, a huge effort has been made to obtain a better series of the sunspot number to characterize the state of the Sun since the invention of the telescope four centuries ago (Hoyt and Schatten 1998; Clette, 2011). However, the current situation is that there are two widely-used sunspot numbers (the International Sunspot Number, ISN, and the Group Sunspot Number, GSN) which differ significantly before 1880. Currently, a series of workshops are being conducted to solve this problem and obtain a consensus series (Cliver, Clette and Svalgaard, 2013).

---

[*] Corresponding author: J.M. Vaquero (jvaquero@unex.es)

Vaquero (2007) noted that an important part of solar observations that were used by Hoyt and Schatten (1998) to construct the GSN series come from observations of the passage of the Sun across the meridian, mainly in the 17th and early 18th centuries.

The most important example of the use of this type of observation is the book published by the Italian astronomer Eustachio Manfredi (1674–1739) who listed the meridian solar observations made during almost one century in the Basilica of San Petronio in Bologna (Manfredi, 1736). More information about the heliometer of San Petronio, a gigantic camera obscura, can be consulted in Heilbron (1999). There are 300 pages in the book by Manfredi (1736) containing more than 4200 observations (made by several scientists during the period 1655–1736) noting the distances of the sun's limbs from the vertex corrected for the penumbra and the apparent diameter of the sun (all given to seconds of arc).

Hoyt and Schatten (1998) used the dates of the observations published by Manfredi (1736). According to them, if we have no information of sunspot on these dates, it is clear that there would be no sunspots because the Sun was really observed on these dates. The number of observations recorded in Manfredi (1736) in each year is shown in Figure 1. The total number of observations is 4204. There are 15 years with no records: 1662, 1664, 1680–1683, 1685–1689, 1692–1694, and 1716. The years with a higher number of records are: 1696 (198 observations), 1697 (193), 1700 (192), 1699 (184), 1701 (182), 1655 (181), 1698 (180), and 1722 (179). Therefore, note that the contribution of these observations to GSN series can be important in some years, specially at the end of the Maunder Minimum (1645-1715, see Eddy, 1976).

Another example of the use of solar meridian observations for the reconstruction of solar activity is the sunspot record that appears in the book *Machina Coelestis* by Johannes Hevelius (1611–1687). Hevelius (1679) listed his daily solar meridian observations from 1653 to 1679 mentioning 19 sunspot groups in one column devoted to comments in his tables. This record was exhaustively studied by Hoyt and Schatten (1995). According to them, these observations are the only known daily listing of solar observations during the early years of the Maunder Minimum and, therefore, they are extremely important for the reconstruction of solar activity during these years.

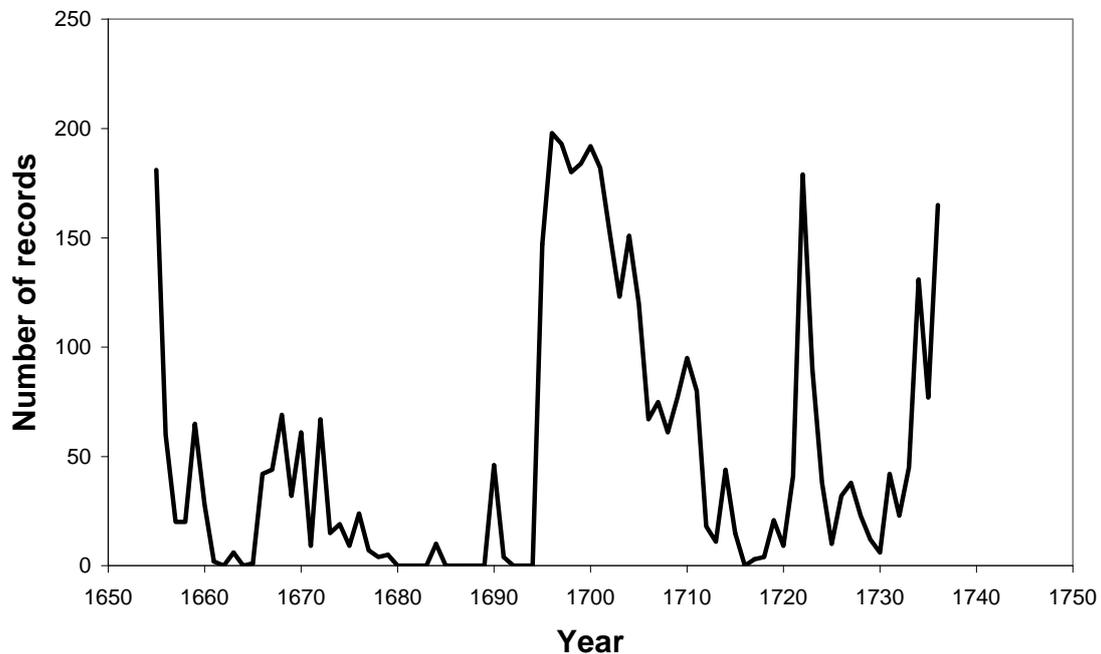

Figure 1. Number of annual solar meridian observations made in the Basilica of San Petronio recorded in Manfredi (1736).

The aim of this paper is to study the sunspot observations made at the Royal Observatory of the Spanish Navy (hereafter, ROA, its Spanish acronym) in the years 1833–1840. The results of these observations can help us to evaluate sunspot observations made in the 17th and 18th centuries during campaigns of meridian solar observations and, therefore, to establish the reliability of these kinds of records for the reconstruction of solar activity.

**2. Meridian Solar Observations at Royal Observatory of Spanish Navy**

The origins of the ROA back to 1753 when the Royal Observatory of Cadiz was founded. In 1798, this Observatory was transferred from Cadiz to the present city of San Fernando (formerly "La Isla de León", 11 km from Cadiz) where it is located today. The ROA was devoted to astrometric work from the beginning, having high-quality instrumentation for this purpose. For a detailed account of the history of this military and scientific centre, see the monographs by Lafuente and Sellés (1988) and González González (1992, 2004).

The meridian solar observations that have been consulted were made with the meridian telescope (D=0.125 m and f=3.05 m) by Thomas Jones, following the meridian telescopes installed in Greenwich and Kensington. It was installed in ROA in 1833, being in use until 1862 (see section IV of the monograph by González González (1992), including an old photograph of the Jones' meridian telescope on page 149).

It is noteworthy that we have both printed and handwritten copies of these observations. The Spanish Government published these observations corresponding to the years 1833, 1834, and 1835 in three volumes (Sánchez Cerquero, 1835, 1836a, 1836b). Moreover, in the historical archive of the ROA (Archivo Histórico del Real Observatorio Astronómico, AHROA) the original manuscripts containing the observations are preserved. In the AHROA, "Astronomía" section, "Observaciones encuadernadas" series, we have consulted the original manuscripts corresponding to the observations in the years 1833 (box 286), 1834 (box 275), 1835 (box 276), 1836 (box 276), 1837 (box 277), 1838 (box 277), 1839 (box 278), and 1840 (box 278).

In these manuscripts there appear, as marginal notes and comments, some descriptions of sunspots. It must be noticed that these marginal notes were not published in the printed observations of 1833, 1834, and 1835. A complete list of these annotations is reproduced in Appendix I containing the date of the report, the name of the observer (all the observers belonged to the staff of ROA), and the original description of sunspots. In total, there are 40 references to sunspots for the years 1835 (4 references), 1836 (14), and 1837 (22). There are no references to sunspots for the years 1833–1834 and 1838–1840. The number of meridian observations is clearly higher than the number of sunspot observations. In particular, In total, there are 1501 solar meridian observations for the years 1833 (121 observations), 1834 (206), 1835 (151), 1836 (163), 1837 (268), 1838 (235), 1839 (139) and 1840 (219).

There are a great variety of descriptions in the forty descriptions retrieved. Some of the descriptions are very detailed as the note of 4 September 1836 ("There is a group notable for its extension in these last days of regular spots that ran in the direction NE-SW met a single spot of extraordinary dimensions, and a significant penumbra. Another group of spots scattered in the same direction, not so great. There is more to the SW, the first one not far from the first limb of the Sun, and a fairly regular spot is near the second limb of the Sun"). However, other descriptions are very vague as the note of the 21 April 1836 ("notable groups of spots remain on the Sun"). Only one of the forty descriptions indicates no sunspots. Thus, the report of June 5, 1836 says: "Spots are not seen". However, Hoyt and Schatten (1998) collected several observations of sunspots this day. For example, H. Schwabe reported five sunspot groups this day.

In spite of this great variety of descriptions, these reports can be considered as a list of notable sunspot groups: more than 60% of the descriptions include words like "remarkable", "considerable", "extraordinarily large" or "large". Moreover, we must stress that a list of the number of sunspots can not be made because there are many qualitative (or oriented to the timing of meridian passage) records.

**3. Discussion**

We would like to evaluate the reliability of the sunspot records of the ROA from the point of view of the reconstruction of past solar activity. Thus, we will make a comparison of these sunspot reports with other solar indices and we will study the frequency histogram of the intervals between the dates of consecutive sunspot reports.

A comparison between the dates of the sunspot reports available in the Appendix I and two well-known solar indices (sunspot area and number) is presented in Figure 2 covering the time-span 1835–1838. The dates of sunspot reports in the meridian solar observations of ROA are represented by green vertical lines. Fortnightly values of sunspot area provided by De la Rue, Stewart, and Loewy (1869) are represented by the red line. The reliability of this data-set in terms of the solar activity indices is discussed in Vaquero, Sánchez-Bajo, and Gallego (2002). Daily values of GSN provided by Hoyt and Schatten (1998) are represented by the blue line.

Unfortunately, no relationship between the timing of reports ROA and the area or number of spots is shown in Figure 2. The astronomers of ROA reported sunspots in times of high and low solar activity. Sometimes, sunspots are reported when there are abrupt peaks in GSN (as occurred around 1836.5) but we can find other examples showing low values of GSN (as occurred around 1835.5).

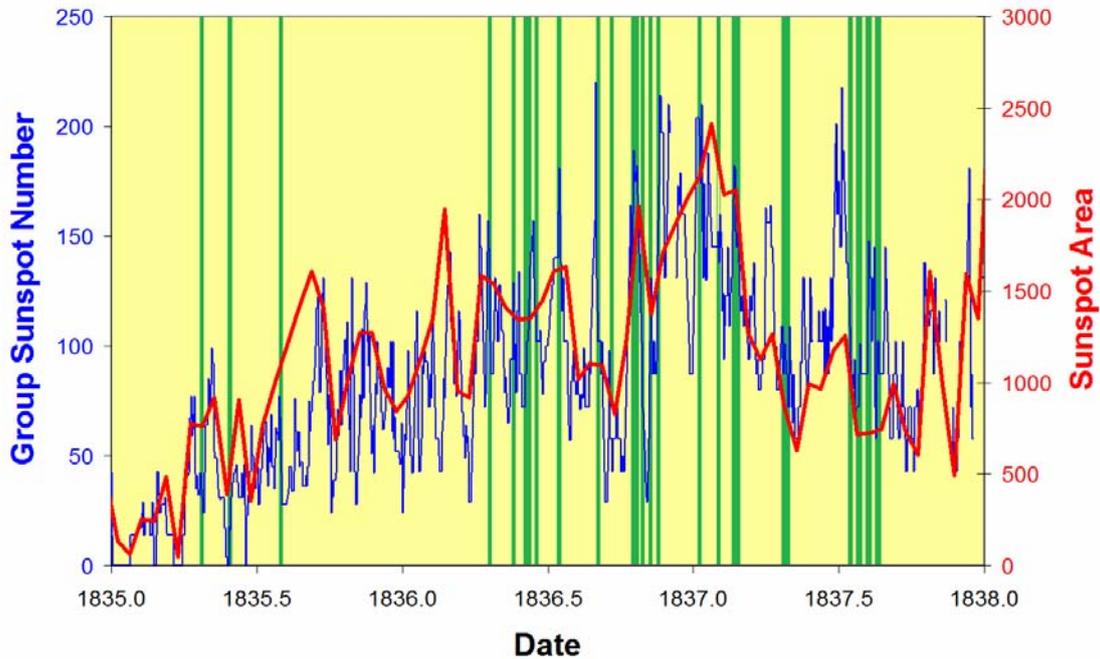

Figure 2. A comparison between sunspot reports in the meridian solar observations of ROA (green) and two solar indices in the same period: (red) fortnightly values of sunspot area provided by De la Rue, Stewart, and Loewy (1869) and (blue) daily values of Group Sunspot Number provided by Hoyt and Schatten (1998).

We would like to verify the statistical significance of the frequency with regard to the different intervals between the dates of consecutive sunspot reports. We have constructed a histogram for this set of observations grouping the intervals between events in bins with 5-day width (Figure 3). If we assume that these reports contain information about notable sunspot groups, we should expect significant peaks around 1-5 days and 25-30 days because great sunspot groups have a long life and because of solar rotation, respectively. We have estimated the statistical significance of our results with a Monte-Carlo simulation. We have generated 200 000 random time series constrained with a similar amount of sunspot reports. The mean value (thick line) and the corresponding values associated to ±1σ (thin line) and ±2σ (dashed line) intervals is shown in Figure 3.

The frequency distribution of the number of cases as a function of the lapse time between sunspot reports is very similar to the average values of the 200 000 random time series constrained with a similar amount of sunspot reports. The only notable difference is found in the value of the bin corresponding to 1-5 days, that surpass the mean values plus 2σ of the 200 000 random time series. This fact supports our comments about these meridian observations as a "list of notable sunspots". Thus, when a remarkable spot was seen, astronomers were observing that spot for a few days. The absence of significant value for the bin of 25-30 days related to solar rotation shows the high degree of randomness in these sunspot records made during solar meridian observations.

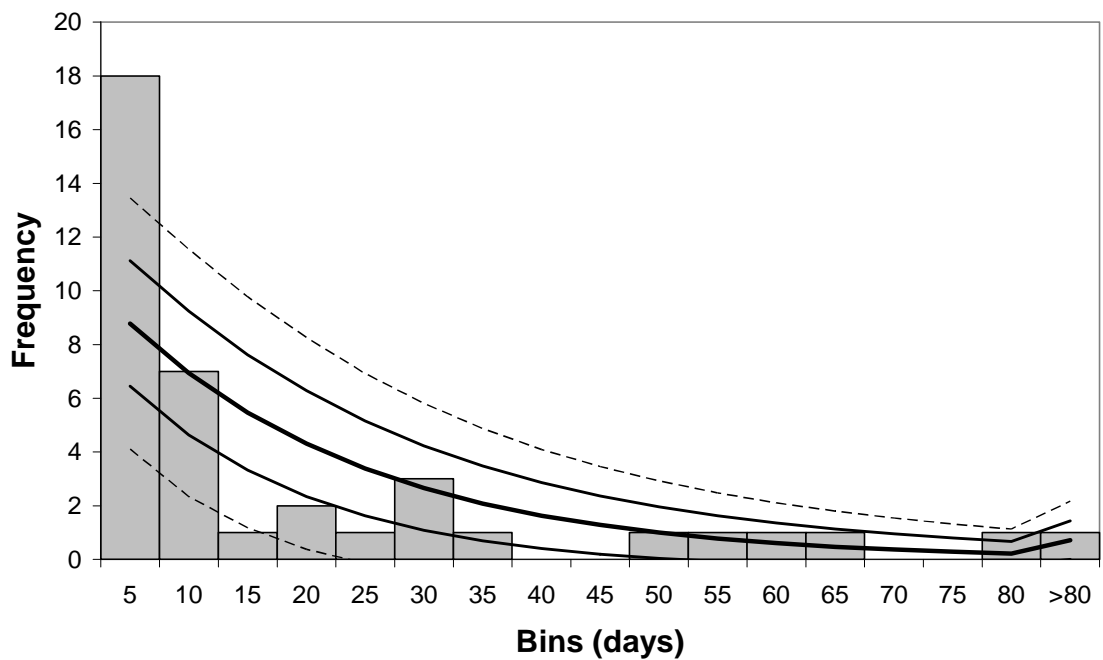

Figure 3. Frequency histograms on the number of cases as a function of the lapse time between sunspot reports. The mean value (thick line) are also shown with the corresponding values associated to ±1σ (thin line) and ±2σ (dashed line) intervals obtained with a Monte-Carlo simulation that have generated 200 000 random time series (constrained with a similar amount of sunspot reports).

Note that this kind of analysis can be problematic if there is a large number of days without observations from our records. Therefore, we have constructed the frequency histogram (Figure 4) of the days elapsed between solar meridian observations made in ROA from 25 April 1835 to 23 August 1837. These dates correspond to the start and end dates of the sunspot record presented in Appendix I. Fortunately, the number of long gaps (over 5 days) between observations is very small (Figure 4).

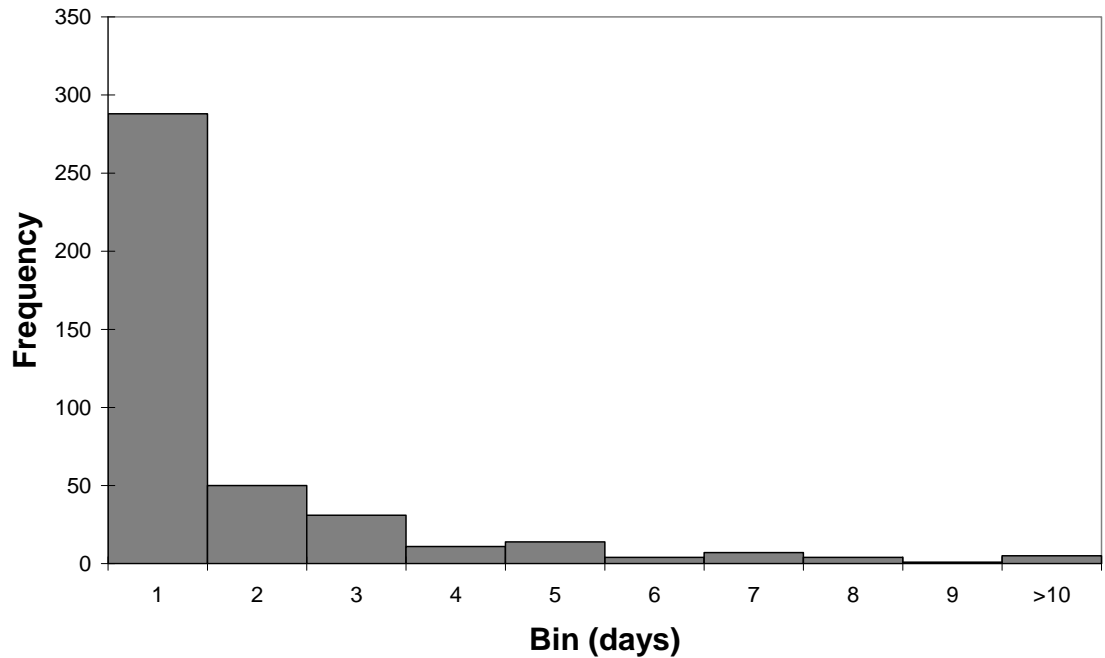

Figure 4. Frequency histogram on the days between solar meridian observations made in the ROA from 25 April 1835 to 23 August 1837.

## 4. Conclusion

We have shown the main features of sunspot records included in the meridian solar observations made at the Royal Observatory of the Spanish Navy in the period 1833–1840:

i) The descriptions given in Appendix I should be considered as a list of notable sunspots.
ii) The sunspot records appear only for the years 1835–1837. Manuscripts of the observations of 1833–1834 and 1838–1840 contain no observations of sunspots.
iii) Records of sunspots only appear in the manuscripts of the observations. These records do not appear in the printed version of the meridian observations.
iv) There is even an incorrect record of no sunspot (although we do not know the origin of this error).
v) There is not a clear relationship between the timing of recorded spots and usual solar activity indices (sunspot area and number).
vi) There are no notable periodicities in these observations.

The sunspots recorded in the manuscripts of the meridian solar observations made in the ROA establish an example of the use of this kind of record to reconstruct the solar activity of the past. According to the results presented in this work, in general, solar meridian observations should be used with extreme caution to evaluate past solar activity.

**Acknowledgements** We wish to acknowledge the assistance provided by the staff of the Royal Observatory of the Spanish Navy, specially to Dr. Francisco José González González (AHROA, San Fernando, Spain). J.M. Vaquero has benefited from the impetus and participation in the Sunspot Number Workshops (http://ssnworkshop.wikia.com/wiki/Home). Support from the Junta de Extremadura (Research Group Grant No. GR10131), Ministerio de Economía y Competitividad of the Spanish Government (AYA2011-25945) and COST Action ES1005 TOSCA is gratefully acknowledged.

**Appendix I: Original texts from manuscript sources in Old Spanish.**

| Date | Observer | Original text [Modern English translation] |
|---|---|---|
| 25 Apr 1835 | Montojo | Muy cerca del 2º limbo del Sol había dos grupos de manchas notables [Two groups of notable spots were very close to the 2nd limb of the Sun.] |
| 30 May 1835 | Montojo | Corría el hilo merid.º una mancha notable cuyo centro antecedía al 2º l [One notable spot, whose centre precedes the 2nd limb, crossed the meridian wire.] |
| 31 May 1835 | Montojo | Antecedía la mancha [al segundo limbo] y estaba al sur del [hilo] ecuatorial [The spot preceded [the second limb] and was south of the equatorial [wire].] |
| 2 Aug 1835 | Montojo | Dos grupos notables de manchas tiene el Sol; el uno muy inmediato al primer margen por encima del hilo ecuat.$^1$: el otro como á un tercio del $^1/_2$ diametro solar del segundo [margen] y bastante debajo del mismo hilo. Es considerable la parte de superficie que cogen sobre el disco del Sol [Two notable groups of spots are on the Sun. The first is close to the first limb above the equatorial wire. The other group is approximately one-third of the solar semi-diameter from the second limb and a good way under the same wire. The portion of spotted area on the Sun's disc is considerable.] |
| 21 Apr 1836 | Montojo | Continúan viéndose grupos notables de manchas en el Sol [Notable sunspot groups continue to be seen.] |
| 21 May 1836 | Montojo | Dos grupos notables de manchas, uno hacia el primer limbo, muy próximo al hilo ecuatorial, y por debajo de él en general; el otro hacia el 2º y mucho más bajas aparentemente respecto del mismo hilo [Two notable groups of spots. One towards the first limb, close to the equatorial wire, and below it in general; the other towards the 2nd limb and apparently much lower relative to that same wire.] |
| 5 Jun 1836 | Montojo | No se ven manchas [No spots are seen.] |
| 10 Jun 1836 | Montojo | Numerosas manchas por debajo del hilo ecuatorial en casi todo el disco: una muy notable por su tamaño (grupo más bien) muy cerca del 2º Limbo por encima del hilo ecuatorial [There are numerous spots below the equatorial wire on nearly all the disc. One very notable for its size (it is a group, rather) is very close to the 2nd limb above the equatorial wire.] |
| 19 Jun 1836 | Montojo | Bastantes manchas por todo el disco [Many spots over the entire disc.] |
| 16 Jul 1836 | Montojo | Bastantes manchas hacia el 2º l. y por todo el disco [Many spots towards the 2nd limb and over the entire disc.] |
| 17 Jul 1836 | Montojo | Bastantes manchas por todo el disco [Many spots over the entire disc.] |
| 4 Sep 1836 | Montojo | Un grupo notable por su extensión en estos días pasados de manchas |
| | | regulares que corría en la dirección NE-SO se ha reunido en una mancha sola de extraordinarias dimensiones y una penumbra considerable. Otro grupo de manchas diseminadas en la misma dirección, no tan grande; hay más al SO (el 1.º no está lejos del 1er l.) y una mancha bastante regular cerca del 2º [A notable group of large, regular spots running NE-SW observed in the past few days has now formed a single spot of extraordinary dimensions and considerable penumbra. Another group of spots scattered along the same direction, not so large. There are more to the SW (the 1st is not far from the 1st limb) and a fairly regular spot near the 2nd limb.] |
| 21 Sep 1836 | Montojo | Un grupo de manchas muy considerable cerca del 2º margen, debe ser el que desapareció hace días por el margen opuesto [A group of very large spots near the 2nd limb. It must be the one that disappeared days ago over the opposite limb.] |
| 18 Oct 1836 | Montojo | Grupos de manchas, algunos de extraordinario tamaño [Groups of spots, some of extraordinary size.] |
| 23 Oct 1836 | Montojo | Muchas manchas por todo el disco [Many spots over the entire disc.] |
| 30 Oct 1836 | Montojo | Se ha formado del 27 acá una mancha considerable próxima hoy al 2º l. [From the 27th until now, a considerable spot has formed which is today close to the 2nd limb.] |
| 9 Nov 1836 | Montojo | Cuatro grupos notables de manchas tiene el Sol, dos hacia el primer limbo y dos hacia el 2º [The Sun has four notable groups of spots, two towards the 1st limb and two towards the 2nd.] |
| 19 Nov 1836 | Montojo | Muchas manchas y de tamaño considerable [Many spots and of considerable size.] |
| 10 Jan 1837 | Montojo | Muchas manchas de consideración casi paralelas al hilo horizontal [Many large spots nearly parallel to the horizontal wire.] |
| 2 Feb 1837 | Martínez | Había ayer sobre el hilo ecuatorial y como unos 4" de tiempo distante del limbo oriental una mancha bastante considerable; hoy se halla como unos 8"5 de tiempo. Su figura es elíptica [Yesterday, a large spot was on the equatorial wire about 4" of time distant from the east limb. Today its position is about 8.5" of time. It is elliptical in shape.] |
| 21 Feb 1837 | Martínez | Varios grupos de machas por debajo del hilo ecuatorial. Había una algo mayor que las demás a 2"2 de tiempo del 2º limbo [Several groups of spots below the equatorial wire. One was slightly larger than the others, at 2.2" of time from the 2nd limb.] |
| 22 Feb 1837 | Martínez | Varias manchas. La que ayer distaba 2"2 del 2º limbo dista hoy 7"95 de tiempo determinada esta distancia por dos apulsos. Cerca del 1er limbo tambien por debajo del hilo ecuatorial había otras dos manchas grandes; el paso por el meridiano de la más cercana al limbo fue a 22$^h$ 22$^m$ 35$^s$10 [Several spots. The spot that yesterday was at 2.2" from the 2nd limb is now at 7.95", this distance of time being determined by two appulses. Close to the 1st limb, also below the equatorial wire, there were two |
| | | other large spots. The meridian transit of that closest to the limb was at $22^h 22^m 35.10^s$] |
| 23 Feb 1837 | Martínez | La mancha cuyo paso meridiano fue ayer $22^h 22^m 35^s10$, pasó hoy a $22^h 26^m 20^s10$ [The spot whose meridian transit yesterday was at $22^h 22^m 35.10^s$, today transited at $22^h 26^m 20.10^s$.] |
| 25 Feb 1837 | Martínez | Dos grandes manchas cerca del $1^{er}$ limbo y por debajo del hilo ecuatorial. Su paso por el meridiano fué á<br>$22^h 33^m 59^s00$ la más próxima al limbo.<br>$22^h 34^m 3^s20$ la más lejana.<br>[Two large spots near the 1st limb and below the equatorial wire. Their meridian transits were at:<br>$22^h 33^m 59.00^s$ the spot closer to the limb;<br>$22^h 34^m 3.20^s$ the farther spot.] |
| 26 Feb 1837 | Martínez | De las dos manchas observadas ayer desapareció la primera. La $2^a$ pasó hoy á $22^h 36^m 45^s98$ observ.$^{ta}$ amp.$^{ta}$ [The first of the two spots observed yesterday has disappeared. The 2nd transited today at $22^h 36^m 45.98^s$.] |
| 27 Feb 1837 | Márquez | Una mancha cerca del $2º$ l. y corriendo el hilo ec.$^l$ pasó á $42^m 32^s6$ h.$^o$ m.$^o$ [A spot near the 2nd limb and running along the equatorial wire transited at $42^m 32.6^s$.] |
| 25 Apr 1837 | Martínez | Un gran grupo de machas por debajo del h.$^o$ ecuat.$^l$ y cerca del $1^{er}$ limbo [A large group of spots below the equatorial wire and near the 1st limb.] |
| 30 Apr 1837 | Martínez | $2^h 29^m 40^s90$ Centro de una gran mancha inmediata al $1^{er}$ limbo y por debajo del hilo ecuatorial [$2^h 29^m 40.90^s$ centre of a large spot next to the 1st limb and below the equatorial wire.] |
| 1 May 1837 | Martínez | El paso del centro de la mancha observada ayer ha sido hoy á 33' 22"55 por tres apulsos [Today the transit of the centre of the spot observed yesterday was at 33' 22.55", by three appulses.] |
| 17 Jul 1837 | Balzolá | Una mancha de consideración dista $3^s5$ del $2º$ limbo [A large spot at $3.5^s$ from the 2nd limb.] |
| 18 Jul 1837 | Balzolá | La mancha dista hoy del $2º$ limbo [The spot today is at $10.6^s$ from the 2nd limb.] |
| 27 Jul 1837 | Balzolá | $8^s 27^s 57^s76$ Centro de dos manchas de una misma ascensión recta y de alguna conside.$^n$ por su tamaño [$8^s 27^s 57.76^s$ centre of two spots of the same right ascension and fairly large in size.] |
| 28 Jul 1837 | Balzolá | Las manchas distan hoy $11^s28$ por cuatro [apulsos] de $1^{er}$. [The spots today are at $11.28^s$ from the 1st limb, by four appulses.] |
| 30 Jul 1837 | Balzolá | Las manchas distan $26^s53$ del $2.^{do}$ l.$^o$, por tres determinaciones [The spots are at $26.53^s$ from the 2nd limb, by three measurements.] |
| 8 Aug 1837 | Balzolá | $9^s 12^s 36^s74$ Mancha de consideración que se acerca más al $2.^{do}$ l. [$9^h 12^m 36.74^s$ Large spot that is closer to the 2nd limb.] |
| 9 Aug 1837 | Balzolá | La mancha observada ayer dista hoy del $1^{er}$ l. $11^s10$. Otra mancha pequeña dista del $2.^{do}$ $4^s07$ [The spot observed yesterday is today at |
| | | 11.10$^s$ from the 1st limb. Another small spot is at 4.07$^s$ from the 2nd limb.] |
| 10 Aug 1837 | Balzolá | 9$^s$ 19$^s$ 56$^s$52 1ª mancha próxima al 1$^{er}$ l. observada los días anteriores. La otra mancha pequeña y cuya distancia al 2.$^{do}$ l. fue determinada ayer por tres apulsos, dá hoy por igual número 10$^s$47 de dis.$^a$ [9$^h$ 19$^m$ 56.52$^s$ the 1st spot observed close to the 1st limb on the previous days. The other small spot, whose distance to the 2nd limb was determined yesterday by three appulses, is today at 10.47$^s$.] |
| 11 Aug 1837 | Balzolá | La primera mancha desapareció. La distancia de la segunda al l. 2.$^{do}$ no he logrado determinarla por la difícil posición en que para ello está la tal mancha [The first spot has disappeared. I was unable to determine the distance between the second spot and the 2nd limb due to the difficult position of this spot for this purpose.] |
| 20 Aug 1837 | Balzolá | 9$^s$ 59$^s$ 22$^s$20 Gran mancha que va delante del 2.$^{do}$ l. [9$^s$ 59$^s$ 22.20$^s$ Large spot that precedes the 2nd limb.] |
| 23 Aug 1837 | Balzolá | La gran mancha dista hoy del 2.$^{do}$ l. 38$^s$27. [The large spot is today at 38.27$^s$ from the 2nd limb.] |